\documentclass{aastex}
\usepackage{amsmath,amssymb}
\usepackage{emulateapj5}

\newcommand{\rmd}{{\rm d}}
\newcommand{\bx}{{\mathbf{x}}}
\newcommand{\BE}{{\mathbb{E}}}
\newcommand{\BR}{{\mathbb{R}}}               
\newcommand{\CA}{{\cal A}}
\newcommand{\CD}{{\cal D}}
\newcommand{\hMpc}{{\ifmmode{h^{-1}{\rm Mpc}}\else{$h^{-1}$Mpc}\fi}}
\newcommand{\average}[1]{\left\langle #1 \right\rangle_\CD}
\newcommand{\baverage}[1]{\left\langle #1 \right\rangle_R}
\newcommand{\inI}{{\bf I}}
\newcommand{\inII}{{\bf II}}
\newcommand{\inIII}{{\bf III}}
\newcommand{\initial}[1]{{#1_{\rm \bf i}}}

\makeatletter
\newenvironment{tablehere}
  {\def\@captype{table}}
  {}
\makeatother

\providecommand{\aanda}{A\&A}
\providecommand{\apj}{ApJ}

\providecommand{\mnras}{MNRAS}
\providecommand{\afz}{Astrophysics}

\begin{document}

%
\title{On the abundance of collapsed objects}
\author{Martin Kerscher$^{1,2}$, Thomas Buchert$^{3,4}$, 
Toshifumi Futamase$^4$}
\altaffiltext{1}{Sektion Physik, Ludwig--Maximilians--Universit{\"a}t, 
Theresienstra{\ss}e 37, 80333 M{\"u}nchen, Germany}
\altaffiltext{2}{Department of Physics and Astronomy, 
The Johns Hopkins University, Baltimore, MD 21218, U.S.A.}
\altaffiltext{3}{Theoretical Astrophysics Division, National Astronomical
Observatory, 2--21--1 Osawa Mitaka, Tokyo 181--8588, Japan}
\altaffiltext{4}{Astronomical Institute, Graduate School of Science, 
Tohoku University, Aoba, Sendai 980--8578, Japan}
\shorttitle{On the abundance of collapsed objects}
\shortauthors{M.\ Kerscher, T.\ Buchert, \& T.\ Futamase} 

\slugcomment{submitted July 17, 2000;  accepted July 31, 2001}

%
\begin{abstract}
The redshift  dependence of the abundance of  collapsed objects places
strong constraints  on cosmological models of  structure formation. We
apply a recently proposed model describing the anisotropic collapse of
inhomogeneous spatial  domains.  Compared with  the spherical top--hat
model,  this  generic  model  leads to  significantly  more  collapsed
objects at  high redshifts: at redshift  one and on the  scale of rich
clusters a factor of 65.
Furthermore,  for a  fixed  normalization of  the initial  fluctuation
spectrum  ($\sigma_8=1$),  we predict  four  times  as much  presently
collapsed  objects on  the  mass--scale of  rich  clusters within  the
standard CDM cosmogony, compared to the spherical collapse.
\end{abstract}


\section{Introduction}

The abundance of galaxy clusters has been studied by
{}\citet{press:luminosity} based on the spherical self--similar
collapse.  If this spherical domain is initially over--dense, one is
led to a model of collapse, featuring a scale--factor (proportional to
the radius of the domain) that, in the course of evolution first
increases with the Hubble flow, then attains a vanishing
time--derivative interpreted as the ``decoupling from the global
Hubble--flow'', and finally shrinks to zero interpreted as ``the
collapse of the spherical region''.  On a given spatial scale the
spherical top--hat model consistently describes an embedded ``small
Friedmann model'', its radius (scale--factor) also obeying Friedmann's
equations but for a mass different from the mass given by the
background density.  This top--hat model is a sub--case of the general
expansion law discussed below.
Deviations from the spherical collapse in a realistic situation have
been studied in numerous works (some of which shall be cited later
on).  Notwithstanding, results from these works have not led to a
simple alternative model for generic {\em anisotropic} collapse of
{\em inhomogeneous} spatial domains, and are mostly considered as
(local) refinements of ``rough but sufficiently accurate'' top--hat
estimations.  In this situation we have recently proposed a model for
a generic collapse as a spin--off from studies of the ``backreaction
problem'' in Newtonian cosmology.  We shall demonstrate that on large
mass--scales this generic model leads to substantially different
results that cannot be regarded as mere refinements of the top--hat
model.


The  generic collapse model  is based  on a  general expansion  law in
Newtonian  cosmology  for  the   dust  model,  obtained  by  averaging
Raychaudhuri's  equation  on   mass--conserving  spatial  domains,  as
summarized below.
Let us denote with $\average{\cdot}$ the spatial averaging in Eulerian
space, e.g., for a spatial tensor field $\CA(\bx,t)=\{A_{ij}(\bx,t)\}$
we simply have the Euclidean volume integral normalized by the volume
of the domain: $\average{\CA}=1/V(t)\int_\CD\rmd^3x\CA(\bx,t)$.
Especially we are interested in the time evolution of the volume
$V=|\CD|$ of the domain $\CD$, modeled by the domain--dependent
scale--factor $a_\CD(t)=\left(V(t)/V(\initial{t})\right)^{1/3}$.
By  averaging Raychaudhuri's  equation  for mass--conserving  domains,
i.e.\      domains      identified      in      Lagrangian      space,
{}\citet{buchert:averaging}   obtained   a   general   expansion   law
describing the evolution of the  volume of a domain, via the evolution
of the scale--factor $a_\CD$:
\begin{equation} 
\label{eq:expansion-law}
3 \frac{{\ddot a}_\CD}{a_\CD} + 4\pi G\average{\varrho} -\Lambda = Q_\CD \;,
\end{equation}
with  Newton's gravitational constant  $G$, the  cosmological constant
$\Lambda$,  and  the  ``backreaction  term'' $Q_\CD$,  which  features
positive  definite fluctuation  terms  in the  parts  of the  velocity
gradient: 
\begin{equation}
\label{eq:Q}
Q_\CD = \frac{2}{3}\average{(\theta -\average{\theta})^2} +
2\average{\omega^2 -  \sigma^2},
\end{equation}
with the  expansion rate $\theta$,
the rate of  vorticity $\omega$, and the rate  of shear $\sigma$.  For
$Q_\CD=0$ this equation equals one  of the Friedmann equations for the
scale--factor  $a(t)=a_\CD(t)$ in a homogeneous  and isotropic universe
with uniform  density $\varrho_{H}=\average{\varrho}$.
The key--difference of the evolution of generic domains compared with
that described by the top--hat model may be summarized as follows: the
collapse of a generic domain is triggered not only by an
over--density,
$\average{\delta}=(\average{\varrho}-\varrho_H)/\varrho_H$, but also
by fluctuations in the velocity gradient encoded in the ``backreaction
term'' $Q_\CD$, most prominently by the averaged expansion and shear
fluctuations.  Consistently, $Q_\CD$ vanishes for spherically
symmetric flows inside spheres.

The   part    still   needed   to   solve    the   general   expansion
law~\eqref{eq:expansion-law}  is a  model  for the  time evolution  of
$Q_\CD$.   {}\citet{buchert:backreaction} (BKS)  calculated $Q_\CD(t)$
for the  spherical and plane  collapse, and also provided  the results
based   on  the  growing   mode  solution   of  the   Eulerian  linear
approximation and on the Lagrangian linear approximation restricted to
Zel'dovich's approximation.   The approximate $Q_\CD(t)$  based on the
``Zel'dovich  approximation'' is  appropriate, if  one wants  to trace
generic initial conditions into the  mildly nonlinear regime.  It is a
powerful       property      of       the       general      expansion
law~\eqref{eq:expansion-law}  combined  with  this backreaction  model
that the  plane--symmetric collapse  {\em and} the  spherical collapse
are included as exact sub--cases.
Important   for   our  work   is   that   the  ``backreaction   term''
$Q_\CD(\inI_R,\inII_R,\inIII_R,t)$   now   only   depends   on   known
time--dependent  functions,  and   the  invariants  $\inI_R,  \inII_R,
\inIII_R$ of the {\em initial} velocity gradient $v_{i,j}$ averaged in
an initially spherical volume with radius $R$ (see BKS).
For a Gaussian random field the volume--averaged invariants $\inI_R,
\inII_R, \inIII_R$ are uncorrelated and the statistical (ensemble)
average of each of them is equal to zero for any domain (BKS).  But
the fluctuations of, e.g.,
$\sigma_{\inI}^2(R)=\BE\left[\inI_R^2\right]$ are non--zero and may be
calculated from the power spectrum.  Specifically, for the
volume--averaged first invariant this is straightforward:
\begin{equation}
\label{eq:initial-I-Pk}
\sigma_{\inI}^2(R) = (2\pi)^{6}
\int_{\BR^3}\rmd^3k\ P(k)\widetilde{W}_R(k)^2 .
\end{equation}
$P(k)$ is the initial power spectrum and $\widetilde{W}_R(k)$ is the
Fourier transform of the top--hat window with radius $R$.  Similarly,
$\sigma_{\inII}(R)$ and $\sigma_{\inIII}(R)$ may be related to the
power spectrum of the density fluctuations (see BKS).  The variance
$\sigma_{\inI}^2(R)$ is equal to the well--known mean square
fluctuations of the {\em initial} density contrast field.

\begin{center}
\begin{tablehere}
\caption{\label{table:sigmas} \footnotesize The r.m.s.\ fluctuations
$\sigma_{\inI}(R)$, $\sigma_{\inII}(R)$, and $\sigma_{\inIII}(R)$ for
an initial domain of radius $R$ are given; the calculation was based
on a standard CDM power spectrum.  To make these numbers more
accessible, also the linearly extrapolated mean fluctuations
$a(t_0)\sigma_{\inI}(R)$ for a domain with scaled radius $a(t_0)R$ at
present time are given.  $M$ is the total mass inside such a domain.}
\vspace{0.1cm}
\noindent
\begin{tabular}{l|lll}
$a(t_0)R\ [\hMpc]$    & 5    & 8    & 12.5    \\
\hline
$a(t_0)\sigma_{\inI}$ & 1.5   & 1.0   & 0.6   \\
$R\ [\hMpc]$     & 0.025 & 0.040 & 0.062 \\
$M\ [10^{15}h^{-1} M_\odot]$ & 0.15 & 0.6 & 2.3 \\
\hline
$\sigma_{\inI}$\ $\times 10^{3}$ 
                 & 7.5   & 5.0   & 3.0   \\
$\sigma_{\inII}$\ $\times 10^{6}$ 
                 & 35    & 16    & 7.4   \\
$\sigma_{\inIII}$\ $\times 10^{9}$ 
                 & 15    & 4.2   & 1.0   \\
\end{tabular}
\end{tablehere}
\end{center}
\vspace{0.1cm}

In the following we assume that the background evolution follows an
Einstein--de~Sitter model with $h=0.5=H_0/100h^{-1}$km/s/Mpc.  The
scale--factor of the background model is normalized to unity at the
initial time, hence $a(\initial{t})=1$ and $a(t_0)=\initial{z}+1$, and
the initial density fluctuations are modeled as a Gaussian random
field with a Cold Dark Matter (CDM) power spectrum
{}\citep{bardeen:gauss}.  We employ the normalization
$a(t_0)\sigma_{\inI}(8\hMpc/a(t_0))=\sigma_8=1$.  For such a situation
the fluctuations $\sigma_{\inI}$, $\sigma_{\inII}$, $\sigma_{\inIII}$
are given in Table~\ref{table:sigmas} for several radii of the initial
domain.
The normalization of the power spectrum enters in a linear, quadratic,
and   cubic   way   into   $\sigma_\inI(R)$,   $\sigma_\inII(R)$   and
$\sigma_\inIII(R)$, respectively.

\section{Generic Collapse}

As an operational definition for the redshift of collapse $z_c$ we use 
the redshift when $a_\CD(z_c)$ approaches zero with $\dot 
a_\CD(z_d)<0$.  Our interest is whether a domain did already collapse 
until present.  Clearly, this will depend on the initial conditions 
and on the radius $R$ of the initial spherical domain, quantified by 
the r.m.s.\ fluctuations of the volume--averaged initial invariants 
$\sigma_\inI(R)$, $\sigma_\inII(R)$ and $\sigma_\inIII(R)$.
To calculate the distribution of $z_c$ we assume that for a fixed
radius $R$ the volume--averaged initial invariants $\inI_R$, $\inII_R$
and $\inIII_R$ are Gaussian random variables with zero mean and
variance $\sigma_\inI(R)$, $\sigma_\inII(R)$ and $\sigma_\inIII(R)$,
respectively.  This is correct for $\inI_R$, but certainly an
approximation for $\inII_R$ and $\inIII_R$ (see the discussion below).
We choose values for $\inI_R$, $\inII_R$ and $\inIII_R$ according to
these probability laws.  Then we solve the general expansion
law~\eqref{eq:expansion-law} numerically for $a_\CD$ as outlined in
BKS starting at a redshift of $\initial{z}=200$ and calculate the
redshift $z_c$, if the domain did collapsed up until $z=0$.  We
iterate this procedure to determine the distribution of
$z_c$. Starting at $\initial{z}=1000$ or $50$ with accordingly
rescaled fluctuations leads to nearly identical results.

For either spherically symmetric ($Q_{\CD}=0$), plane--symmetric (see
Eq.~(44) in BKS), or generic initial conditions (see Eq.~(42) in BKS), we
estimate the probability $p(z_c)\rmd z_c$ that a domain collapses at
$z_c$, using the Monte--Carlo procedure outlined above.  Hence,
$\int_{-\infty}^0\rmd z_c\ p(z_c)$ is the fraction of domains which
collapsed until present, and $F(z_c)=\int_{-\infty}^{z_c}\rmd z\ p(z)$
is the fraction of the domains which already collapsed up to a
redshift of $z_c$.
\begin{center}
\epsscale{0.94} 
\plotone{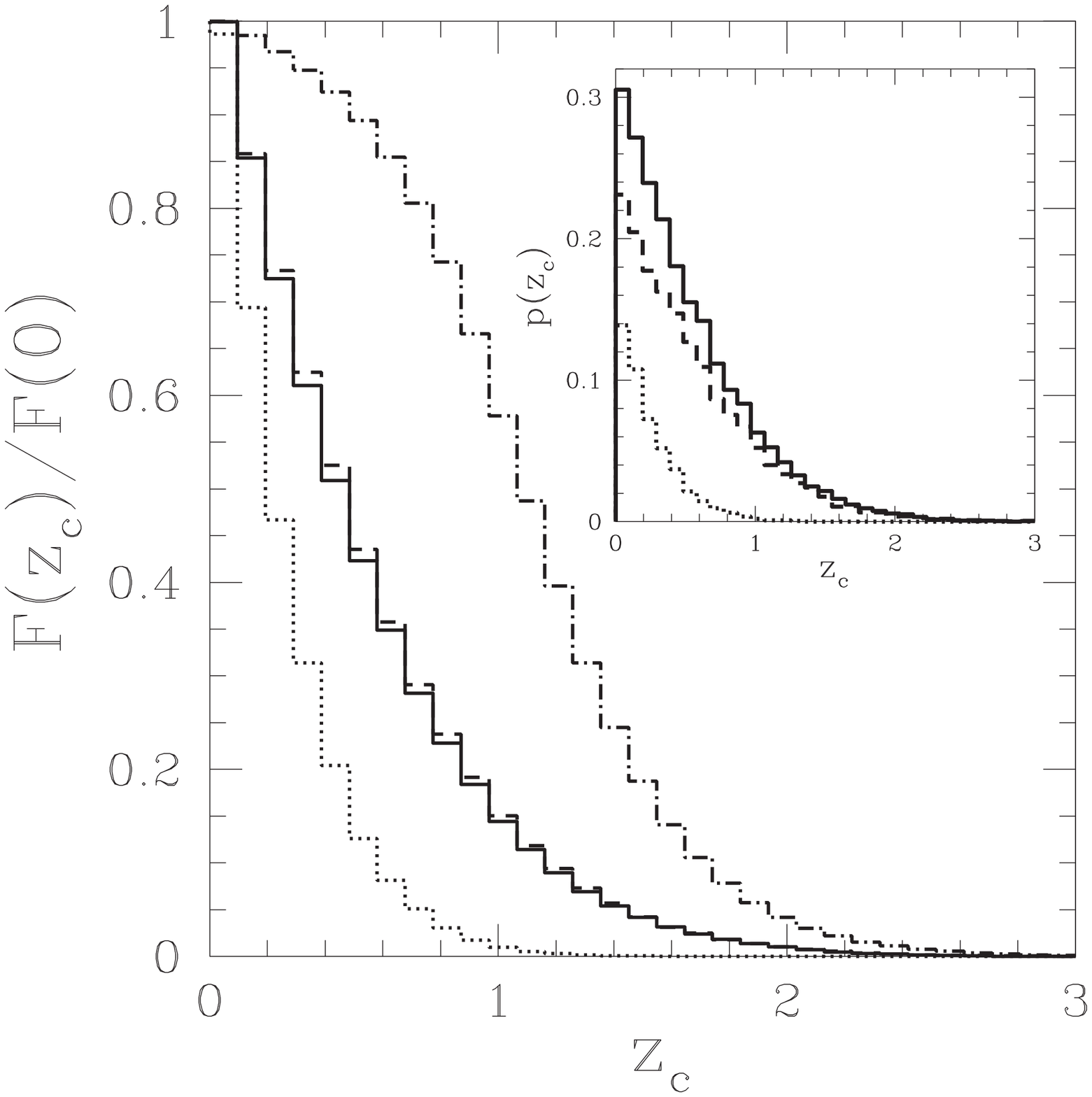}
\figcaption[]{\label{fig:collapsehist} The fraction $F(z_c)/F(0)$ of
collapsed regions with comoving ``radius'' of 8\hMpc\ is shown:
generic initial conditions (solid line), the plane collapse (dashed
line), the spherical collapse (dotted line), and the generic collapse
for restricted initial conditions with $\inI_R\le-1.69$ (dashed dotted
line). In the inset plot $p(z_c)$ is shown.}
\end{center}
In Fig.~\ref{fig:collapsehist} both $p(z_c)$ and $F(z_c)/F(0)$ are
shown for a domain with initial radius of 0.04\hMpc, a comoving
``radius'' of 8\hMpc\ corresponding to a mass--scale of rich clusters.
The most striking feature is that the generic model leads to
significantly more collapsed regions at high redshift in comparison to
the spherical model.  Moreover, only 5\% of the domains collapsed in
the spherical model, whereas in the plane--symmetric case 16\%, and in
the generic case already 20\% of the domains collapsed.
{}\citet{richstone:lower} studied $F(z_c)/F(0)$ using the spherical
model (see {}\citealt{bartelmann:timescales} for consistent initial
conditions).  Considering domains with an initial radius
$5\hMpc/a(t_0)$ we find the same result as
{}\citet{bartelmann:timescales} for the $F(z_c)/F(0)$ in the spherical
model.
Initially over--dense domains with $-\inI_R=\baverage{\delta}\ge1.69$
will always collapse in the top--hat model.  In an Einstein--de~Sitter
universe the top--hat model predicts that domains collapsed only
recently (Fig.~\ref{fig:collapsehist}). As already mentioned,
$\inII_R$ and $\inIII_R$ are stochastically independent of $\inI_R$
for a Gaussian random field (see BKS) and we may investigate the
generic case under the condition that $\baverage{\delta}\ge1.69$.  Now
the majority of these initially over--dense domains already collapsed
before a redshift of one.

To investigate how strong our results depend on the assumption of a
Gaussian law for $\sigma_\inII(R)$ and $\sigma_\inIII(R)$ we repeat
our calculations keeping $\sigma_\inI(R)$ fixed but doubling (halving)
$\sigma_\inII(R)$ and $\sigma_\inIII(R)$.  The increased fluctuations
mimick the effect of a pronounced tail of the distribution and
consequently lead to even stronger deviations from the top--hat model.
For reduced fluctuations the results stay in--between the generic
model and the plane collapse.  In both cases the redshift dependence
of the normalized $F(z_c)/F(0)$ shows nearly no difference compared to
the generic case given in Fig~\ref{fig:collapsehist}.

Up to now we focused on the collapse of domains with an initial radius
$R=8\hMpc/(\initial{z}+1)$ containing the mass
$0.6\times10^{15}h^{-1}M_\odot$.  In the following we will show how
this accelerated collapse depends on the mass--scale.  Using our
Monte--Carlo procedure outlined above, we calculate the fraction
$F_{\text{generic}}(z_c)/F_{\text{spherical}}(z_c)$ for a given
mass--scale quantifying the deviation of the generic from the
spherical collapse.
\begin{center}
\epsscale{0.94}
\plotone{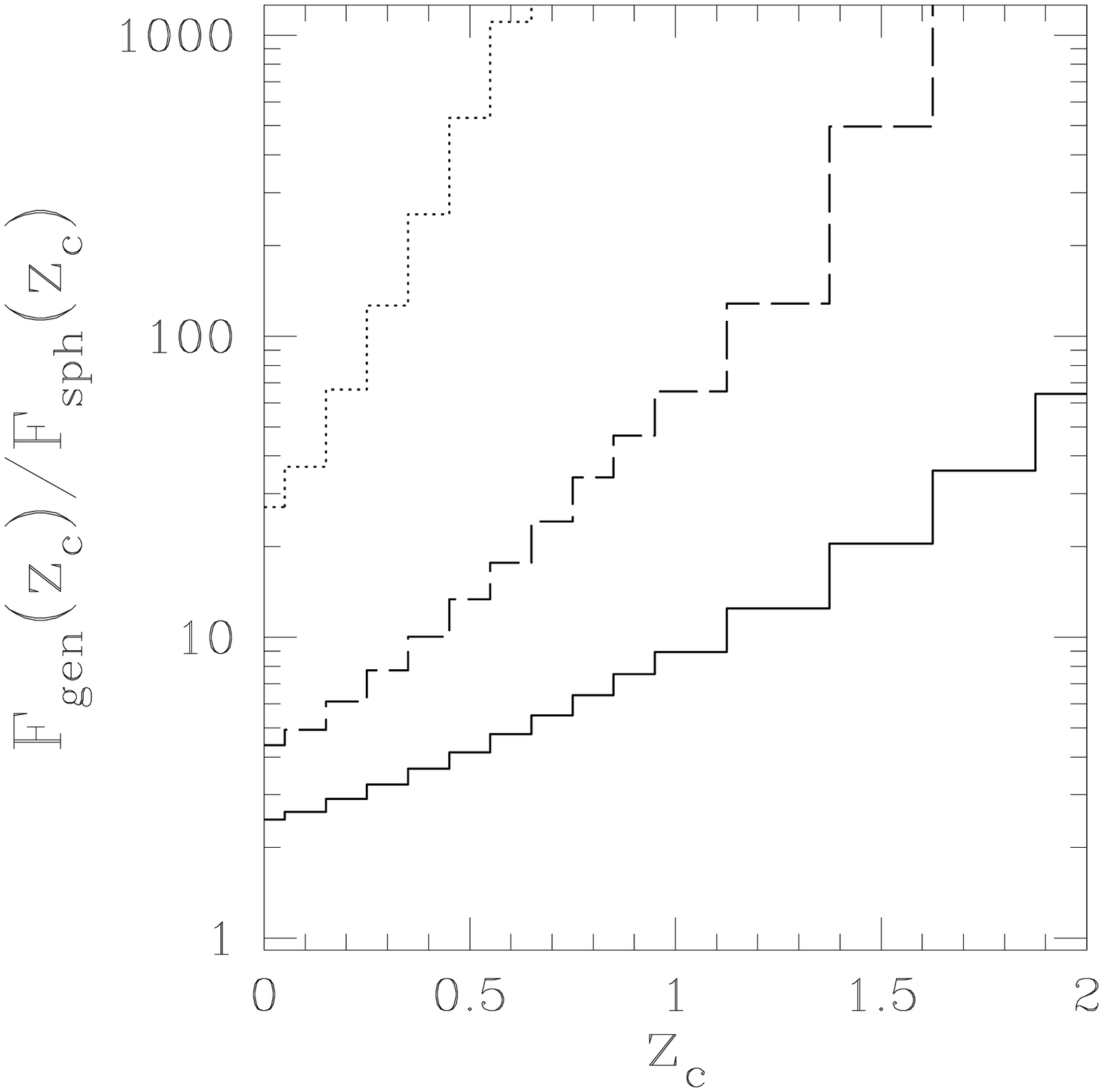} 
\figcaption[]{\label{fig:massdepend}
The ratio $F_{\text{gen}}(z_c)/F_{\text{sph}}(z_c)$ is plotted against
the redshift of collapse $z_c$ for domains with a comoving size of
5\hMpc\ (solid line), 8\hMpc\ (dashed line), and 12.5\hMpc\ (dotted
line).}
\end{center}
In Fig.~\ref{fig:massdepend} we see that at redshift $z=0$ the
spherical model under--predicts the abundance of galaxy clusters on
the mass--scales considered.  For smaller mass--scales
($\le10^{14}h^{-1}M_\odot$) this discrepancy gradually
vanishes. However at higher redshifts and for larger mass--scales we
predict significantly more collapsed domains, e.g.\ a factor of 65 at
$z=1$ and at a mass--scale of rich clusters
($6\times10^{14}h^{-1}M_\odot$).
We want to emphasize that the ``Zel'dovich approximation'' entering 
our calculations in the generic case becomes increasingly more 
accurate for larger domains.
As discussed by {}\citet{kitayama:semianalytic} the SCDM model with
$\sigma_8=1$ over--predicts the present abundance of galaxy
clusters. Our main interest is in the comparison of the spherical
collapse model with the generic anisotropic collapse model, using SCDM
as a well studied reference model.

\section{Discussion}
\label{sect:discussion}

We examined the collapse of initially spherical domains embedded into
a generic inhomogeneous cosmology.  We combined the general expansion
law for such domains with a ``backreaction model'' based on the
``Zel'dovich approximation'' to obtain explicit solutions for the
scale--factor of such domains.  Although we approximated the
backreaction term, the resulting model is more general than the volume
deformation based on the averaged ``Zel'dovich approximation''. It
still contains the spherical top--hat model as exact sub--case and is
also exact in the orthogonal case of plane--symmetric collapse.
Assuming Gaussian initial conditions of a standard Cold Dark Matter
cosmogony we calculated the abundance of collapsed domains.  For a
normalization of the initial fluctuation spectrum with $\sigma_8=1$,
collapsed domains on the mass--scale of rich clusters are more abundant
by a factor of four.  Clearly, these results depend on the adopted
normalization and background model.  However, nearly independent of
the normalization is the disagreement between the top--hat model's
prediction and that of the generic anisotropic collapse model
concerning the abundance of collapsed objects at high redshift:
we obtained 65 times more collapsed objects at a redshift of one on 
the scale of rich clusters. 
Compared to the spherical model, which focuses only on the mass
over--density, our generic collapse model considers also the
fluctuations in the expansion rate as well as the effect of shear
fields on the collapse (see Eq.~\eqref{eq:Q}).  Our results indicate
that on large mass--scales the shear fields, caused by internal and
external mass concentrations, accelerate the collapse.

As a rule, our model mimicks that of N--body results at least down to
the scale where the local (truncated) Lagrangian approximation
reproduces the results of N--body runs {}\citep{weiss:optimizing}.
Only recently conducted large N--body simulations were able to provide
enough dynamic range for the determination of the abundance of
collapsed domains on the high mass--end
{}\citep{jenkins:mass,bode:evolution}. Similar to our results both
studies show that the spherical model under--predicts the abundance of
collapsed objects at high mass--scales. The abundance of high--mass
objects drops significantly with increasing redshift. Hence the
statistical significance of these results from N--body simulation is
limited by the small number of massive halos.  Nonetheless
{}\citet{bode:evolution} observe a similar growing deviation from the
predictions of the spherical model with increasing redshift.
Clearly, the prediction of an increase in the abundance of collapsed
objects by a factor of 65 at $z=1$ needs to be verified against future
simulations.
If we normalize the model to the observed cluster abundance with
$\sigma_8=0.6$ {}\citep{viana:cluster}, the relative difference
between the spherical and the generic collapse becomes even stronger
(see also {}\citealt{governato:properties}).

{}\citet{sheth:ellipsoidal} estimated the mass function assuming an
anisotropic ellipsoidal collapse, where the time evolution of the
half--axes is determined by the ``Zel'dovich approximation''.  Our
model may be viewed as a generalization of this ansatz, where the
domain is not restricted to be ellipsoidal throughout its evolution.
Several other approaches are also based on the ``Zel'dovich
approximation'', sometimes the third--order Lagrangian perturbation
approximation is employed. In these ``local'' approaches the collapse
of a domain is associated with a diverging {\em local} matter density
(see e.g.\ {}\citealt{bartelmann:timescales},
{}\citealt{monaco:lagrangianI,monaco:lagrangianII},
{}\citealt{lee:cosmological}, and {}\citealt{engineer:nonlinear}).
The redshift of collapse is calculated from the distribution of the
eigenvalues of the deformation tensor {}\citep{doroshkevich:gauss}.
The eigenvalues of the deformation tensor are determined locally at
each point in space from a smoothed density field, where the smoothing
scale determines the mass--scale.
Compared to this pointwise treatment, our model is based on the time
evolution of a finite domain. The invariants, averaged over the volume
of the domain, determine the time--scales of the collapse. Therefore,
our approach explicitly takes the spatial correlations of the initial
velocity and density fields into account.

We calculated the fraction of collapsed domains for a given
mass--scale, regardless whether these domains are included in a larger
collapsed region.  This is the cloud--in--cloud problem as tackled by
{}\citet{bond:excursion} and {}\citet{jedamzik:cloud} for the
spherical collapse (see also {}\citet{yano:limitations} taking into
account spatial correlations).  Due to the drop of the mass--function
at the large--mass end, the cloud--in--cloud problem mainly affects
the smaller mass--scales. {}\citet{lee:cosmological} showed in their
calculations of the mass--function that considering the
cloud--in--cloud problem mainly changes the normalization, but not the
shape.  Hence, for the mass--scales considered in this {\em Letter} we
do not expect a significant impact on the time--scales of the collapse.

Still, the question remains, how the abundance of collapsed domains in
this generic model may be related to the abundance of observed
clusters.  In the spherical model one can use the virial theorem to
estimate the over--density where the collapse stops and the system
``virializes''.  However, for small generic domains embedded into a
fluctuating background and admitting non--negligible generic shear
fields on the size of the domain, neither the spherical assumption nor
the virial theorem for isolated systems {}\citep{chandrasekhar:tensor}
holds.  

Both, the spherical and the generic dust model lack forces that could
compensate the collapse.  Besides baryonic physics
{}\citep{gunn:infall} it would be mandatory to include multistream
forces opposing the collapse {}\citep{buchert:multistream}.  While a
spherical collapse is never realized, the generic model includes the
effect of shear fields and fluctuations of the environment of the
domain to describe a realistic collapse situation.

The results presented in this {\em Letter} have an obvious and
important implication for ongoing high--redshift cluster survey
projects using large telescopes such as Subaru and VLT.  Our study
shows the quantitative importance of modeling a realistic collapse
situation for an accurate theoretical prediction of the
mass--function.

\acknowledgements  
We would  like to thank  Claus Beisbart, Alvaro  Dom{\'\i}nguez, Tetsu
Kitayama, Yasushi  Suto, and  Paolo Tozzi for  interesting discussions
and helpful comments on the manuscript.
MK acknowledges  support from the NSF  grant AST 9802980  and from the
{\em Sonderforschungsbereich 375 f{\"u}r Astroteilchenphysik der DFG}.
TB  acknowledges   generous  support  by   the  National  Astronomical
Observatory,  Tokyo, as well  as hospitality  at Tohoku  University in
Sendai, Japan.


\end{document}